# Microwave Power Standard using Cold Atoms

Dave Paulusse, Nelson Rowell, and Alain Michaud
*National Research Council of Canada, INMS,
M-36, 1200 Montreal Road, Ottawa, ON, Canada, K1A 0R6*
*Email:* Alain.Michaud@nrc.ca

We discuss how the observation of Rabi flopping oscillations in a laser cooled atomic sample could be used as a microwave power standard. The rubidium atoms are first trapped in a standard MOT, then optically pumped, and dropped. As they enter the interaction region, a resonant pulsed microwave field is applied. Following the interaction lasting up to 10 ms, a probe laser beam is turned on and the fluorescence measures the population inversion.

Figure 1(a) shows the oscillation for a typical field. The microwave field is resonant with the hyperfine structure of $Rb^{87}$ atoms. The experiment was repeated many times, varying the microwave pulse length. The observed periodic function, with as many as 15 cycles, allows us to determine the frequency precisely. Figure 1(b) shows this frequency as a function of the field amplitude. Any variation from linearity is less than 0.2 %.

Knowledge of the oscillator strength of the transition and the radiation pattern of the waveguide would allow this method to be used as an absolute standard for microwave power measurement.

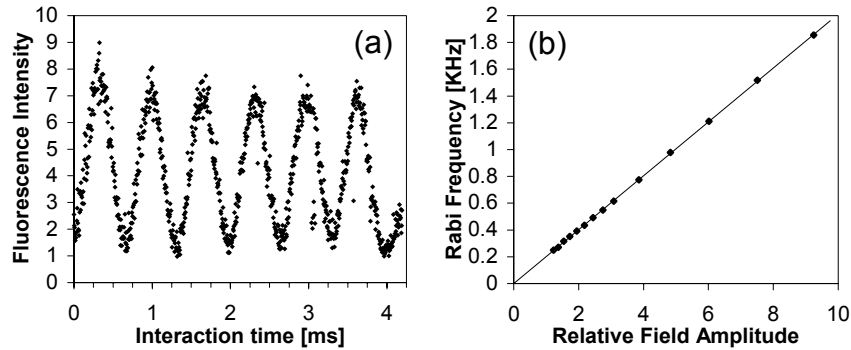

Figure 1(a) Flopping oscillation for a typical microwave power (at 6.8 GHz). (b) Dependence of the flopping frequency as a function of microwave field amplitude.